# Identifying sparse and dense sub-graphs in large graphs with a fast algorithm

Vincenzo Fioriti[1(a)] and Marta Chinnici[1(b)]

[1] *ENEA CR Casaccia - Via Anguillarese 301, 00123 Rome, Italy*

**Abstract** – Identifying the nodes of small sub-graphs with no *a priori* information is a hard problem. In this work, we want to find each node of a sparse sub-graph embedded in both dynamic and static background graphs, of larger average degree. We show that exploiting the summability over several background realizations of the Estrada-Benzi communicability and the Krylov approximation of the matrix exponential, it is possible to recover the sub-graph with a fast algorithm with computational complexity $O(N\,n)$. Relaxing the problem to complete sub-graphs, the same performance is obtained with a single background. The worst case complexity for the single background is $O(n\,log(n))$.

**Introduction.** – Identifying nodes of a small sub-graph embedded in a much larger graph with no *a priori* information is a hard problem. However, many industrial, financial, pharmacological, gene expression, protein structure prediction, evolutionary tree, security and big data applications require to tackle this issue and to do it efficiently, as the computational effort is a major concern. If the sub-graph is a clique (a complete graph) [1,2,3] the problem is somewhat easier; for the general case, interesting results have been claimed using the new emerging methodology of the Signal Processing on Graphs (SPoG) in simulations and real graphs [4]. Although the non-euclidean nature of data represented by graphs complicates the mathematical treatment, classical signal processing would be an important tool if it could be applied to graphs. The analogy is clear: considering "noise" a background graph and "signal" an embedded target sub-graph, the noise filtering would reveal the signal (the sub-graph). Note that the problem here is to identify a particular sparse sub-graph, rather than to analyse communities [1].

In this work, we want to identify each node of a sub-graph embedded in a synthetic background environment of larger average degree. To do so, firstly we consider a source producing several Erdos-Renyi graph (ER) and apply our filtering algorithm to the ER ensemble.

Relaxing the problem to complete sub-graphs (cliques), we recover the signal using only one background (Erdos-Renyi, Barabasi-Albert and Strogatz-Watts) graphs.

**Signal processing on graphs.** – A simple graph $G(n, E)$ is a collection of relations among $n$ objects called nodes or vertices, linked by E edges, indicating that some sort of interaction exists between a pair of nodes. A symmetric matrix called adjacency matrix **A** represents mathematically the interactions: $a_{ij}$ entries are 1 if an edge links node $i$ to node $j$, 0 otherwise. If no direction is specified on the edge, the graph is undirected. Usually, if the graph is derived from technological processes or natural phenomena it is called a network.

Graph properties are described by several parameters such as degree distribution, betweenness, closeness, clustering, eigenvalues spectra and so on; all these and many other techniques involve heavy computations. A complete graph is a graph with all-to-all edges; its density (roughly the ratio between edges and nodes) is 1. If the density is close to the minimal number of edges, the graph is sparse. Sub-graphs are graphs whose nodes are subsets of the larger graph and thus are contained in it.

In the SPoG framework, signals on graphs are defined as discrete values related to nodes, that is, vectors in $C^N$ (see Figure 1) [5]. Of course, the classical signal processing cannot be translated immediately to the graph signal domain. Nevertheless, a number of useful signal processing tools on graphs have been devised, namely the Fourier transform on graph, low-pass and high-pass filters on graphs.

This new point of view has been implemented to develop metrics to reveal anomalies in large graphs, such as a small set of nodes that do not fit the typical behaviour observable in the graph [5]. The analogous problem in the time domain is the detection of a time series embedded in noise. Also the difficulties are the same: if the amplitude of the noise is much larger than the signal amplitude, it is hard, if not impossible, to recover the signal.

Since we consider undirected graphs, the only differences between the target sub-graph and the background graph are the degrees and the infra/inter connection topology. Therefore, if the background is composed by many hubs (nodes with many edges), a high average degree and a connection topology similar to the target, the identification is not trivial. Even the clique decision problem (decide whether a graph contains or not a certain clique) would be NP-hard [6,7].

[a]E-mail: vincenzo.fioriti@enea.it
[b]Present address: ENEA CR Casaccia, - Via Anguillarese 301, 00123 Rome, Italy.

The general problem we are interested in can be stated as follows: *find the nodes of sparse or dense sub-graphs (with a fixed number of nodes), embedded in larger synthetic (possibly real), static or dynamic background graphs, with no a priori knowledge neither of the target nor of the background, other than the target differs somehow from the background*. Differences might be minimal or relevant, but in any case they are not known in advance, while the simultaneous presence of several similar targets or the variation in the number of nodes are not taken into consideration here. In the dynamic case, the background graph is allowed to change the edges while the number of node is fixed.

*Current state of the art.* Given the above definition, an accurate analysis has been provided by Miller [6,7]. Miller uses the dynamic modularity matrix **B**($k$):

$$\mathbf{B}(k) = \mathbf{A}(k) - \mathbf{d}_i(k)\mathbf{d}(k)_j^T / 2|E(k)| \quad (1)$$

where **A** is the observed adjacency matrix, $\mathbf{d}_i\mathbf{d}_j$ are the products of the degrees, $E$ is the number of edges, and $k = 1, 2, ... N$ indicates the time the matrix is evaluated. If **B**($k$) (or **B** in the static case) is projected on a lower dimensional space by an eigendecomposition, the $L1$ norm of the eigenvectors may unveil an anomaly, as the norms of the eigenvectors related to the target are smaller than the others.

After one anomalous node has been detected, the other nodes are identified by a *K*-means clustering algorithm. Starting from the anomalous node, the *K*-means finds two node clusters: the larger is the noise, the smaller is recognised as the target.

Extensive Monte Carlo simulations have been conducted to validate this procedure [6]. Firstly, a target sub-graph of 12 nodes has been embedded in a synthetic static background of 1024 nodes (average degree 11.5). Varying the density of the target (number of edges related to the number of nodes $n$) the identification performance has shown significant results, beginning with the 80% density to a complete identification at 100% density level. In this last case, the average degree of the target was slightly less than the background average degree 11.5. In a second simulation on synthetic dynamic graphs, since (1) may be written as:

$$\widetilde{\mathbf{B}}(k) = \sum_l \mathbf{B}(k-l)c_l, \text{ with } l = 0, ... L-1 \quad (2)$$

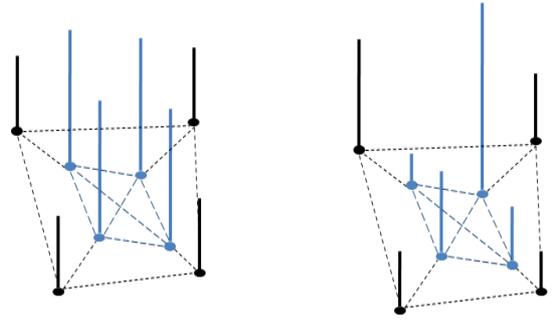

Fig. 1: Signals on a graphs are defined as discrete values situated on the nodes. Left: a deterministic signal; right: a random signal. The blue dots indicate a four nodes clique.

that is, an averaging of the modularity matrix over $N$ realizations of the backgrounds according to the coefficients $c_l$ was implemented in order to filter the noise. In the time-frequency domain, assuming the signal is slowly varying with respect to noise, the averaging reinforces the signal samples while averages-out the nose samples. Something similar happens here with the L1 eigenvectors norm.

The procedure increases the detection power (in the sense of the decision problem) with respect to the static case, while the identification performance (recovering all the nodes) decreases.

Similar simulations were conducted with a real sub-graph (20 nodes, average degree 2.5, density 0.13, the largest hub has degree 8) obtained from a social network. The sub-graph was embedded in a synthetic background of average degree 4 and again the detection was almost sure, despite the lack of intra-connectivity of the target.

However, although interesting, this approach suffers from the amount of calculations required.

**Identification algorithm.** – As first step we discuss the main tools of our procedure, then we will present the algorithm and some applications to synthetic graphs or to real technological networks, along with the simulation results.

*The total communicability.* The Estrada-Benzi total communicability parameter (EBTC) is the main tool of this work, deriving from the Estrada sub-graph centrality [8,9]. The EBTC parameter was devised to tackle the problem of massive computations required by large graphs, the major drawback of the original Estrada sub-graph centrality [9] as well as many other centrality parameters. The Estrada communicability is a generalization of the concept of shortest path including walks. Since walks are sensitive to structural anomalies and bottlenecks, the Estrada parameter looks like an ideal tool for identification purposes and to quantify the information flow among nodes. In fact, it distinguishes finer structures as nodes with larger communicability respect to the rest of the graph, therefore it is suitable to *identify a sub-graph non consistent with the background*. The Estrada sub-graph centrality (**SC**):

$$\mathbf{SC} = e^{\mathbf{A}}|_{ii} \qquad i = 1, .. n \qquad (3)$$

that is the diagonal elements of the adjacency matrix exponential, resulting from the embedding of the target in the background. A large $SC_i$ indicates that the node $i = 1, ... n$ is relevant to the communications flowing in the graph and most of all, characterize the role of a node in a sub-graph. Benzi [8] extends these ideas to the total communicability:

$$\mathbf{EBTC} = \sum_j e^{\mathbf{A}}|_{ij} \qquad j = 1, ... n \qquad (4)$$

simply summing the row entries of $e^{\mathbf{A}}$. Since diagonal values are usually large, (3) and (4) produce similar results depending on the spectral gap (difference between the first and the second largest eigenvalue), at least for ER graphs.

It must be stressed that extracting the diagonal entries of $e^{\mathbf{A}}$ would require much computational time, but if we rely only on the row sums, the Krylov algorithm [10] allows to get directly these sums calculating $e^{\mathbf{A}} \mathbf{1}$ with a linear complexity $O(n)$. As a matter of fact, in our experiments the Krylov approximation proved to be about a hundred times faster than the traditional methods, although speed depends strongly on the graph topology and higher speeds could be reached [8]. Therefore we prefer to trade off accuracy for speed.

Once the EBTC for each node has been calculated, we will look for the largest 20 EBTC values. This can be done efficiently with the Introsort sorting algorithm, whose worst case complexity for very large $n$ is $O(n \log(n))$ [13].

*Multiple backgrounds scenario.* Our initial scenario consists of multiple ER background graphs ranging from an average degree 2 to 6 (1024 - 5000 nodes), and a target sub-graph of average degree 2.1 (20 nodes, hubs of degree 4, density 0.11) as in Figure 2. The sub-graph was taken from Miller and modified in order to reduce the average degree and the degree of the one hub. As pointed out also in [6], the target of Figure 2 is a severe test for an identification algorithm, although ER graphs have a low density.

The target nodes were distributed uniformly at random in the graph for each simulation run. The embedded sub-graph topology does not change, although the degree of the target nodes may increase depending on the links with the background nodes. Up to $N = 40$ different ER backgrounds were produced during each run to simulate an evolving graph.

The Estrada-Benzi total communicability (EBTC) was calculated for each node, directly from the adjacency matrix $\mathbf{A}$ resulting after each embedding, then results were summed. Therefore the EBTC for the $i$-th node over $N$ backgrounds is:

$$EBTC(i) = EBTC^i_1 + EBTC^i_2 + ... + EBTC^i_N \qquad (5)$$

This sum reminds of the Miller filtering, but is faster, simpler and nevertheless, effective. Above all, no averaging takes place here, just a sum and a sorting. Finally, candidate nodes were selected choosing those ranking in the first 20 largest *EBTC* values, as in Figure 3.

As can be seen from Table 1, after testing $N = 40$ ER backgrounds (average degree ~2.1) with the sub-graph described above, the algorithm identifies correctly the 97% of the 20 nodes (averaging over several simulation runs). The 100% rate was reached in half of the runs. The effect of summability is clearly visible in the first three columns of Table 1. Using only two backgrounds reduces the identification rate to 50%, but increasing to 20 backgrounds restores the rate to 92%, although the number of nodes is 9000 (third column).

Increasing the size of the backgrounds from 2000 to 4000 nodes and the average degrees to 3 and 5, similar rates are obtained. These good results are due partly to the random nature of the Erdos-Renyi graphs and partly to the summability of the Estrada-Benzi total communicability.

Increasing further the background average degree, the identification rate drops rapidly, thus we relax the problem to the complete sub-graph. A clique (average degree 19) is embedded in small-world (SW) background (average degree 40, 1024 nodes). Despite the number of backgrounds to be used in (5) was dropped to 2, the identification rate rises high (see Table 2). The summability effect is relevant when the rates are initially low, otherwise, if the identification is easy, the rates are immediately high and therefore the summability contribution is negligible.

The $N$ different backgrounds should mimic a dynamically evolving graph. However, in a real environment the background does not necessarily change its edge connections completely and anyway temporal correlations between consecutive backgrounds would be fortuitous. It is likely that in the real world only a partial transformation occurs. Our simulations of the partial transformation show that the identification rates (not reported here) are not encouraging. Moreover, it is not easy to made available to the analysis an evolving graph, since it would mean a heavy monitoring task of a real process.

Nonetheless, the multiple scenario may find application to the temporal extensions of centrality parameters and metrics based on temporal shortest path, such as infection immunisation modelling and malware spreading [11].

Finally we note that, as a matter of principle, the general problem could be stated as the search for the maximum density sub-graph, since all ER backgrounds used here have density lower than 0.08 to be compared with the target 0.11. However, the maximum density sub-graph problem may be solved in a polynomial time $O(n^k)$, $k < 1$ only if $n$ is not fixed, otherwise the problem is NP-hard [12].

To face these issues, it would be better to take a single snapshot of the process, that is, to use a single background.

Table 1: Multiple backgrounds, *sparse* target (with average degree 2.10). Note in the first three columns the effect of the multiple backgrounds on the identification rate.

| Background[a] | ER | ER | ER | ER | ER | ER |
|---|---|---|---|---|---|---|
| nodes | 1024 | 1024 | 9000 | 2000 | 3000 | 4000 |
| av. deg. bgr. | 2 | 2 | 2.1 | 2.9 | 3.9 | 4.9 |
| nr. backgr. | 40 | 2 | 20 | 40 | 40 | 40 |
| % id. rate | 97% | 50% | 92% | 96.5% | 95% | 82% |

[a] All backgrounds have density less than 0.05.

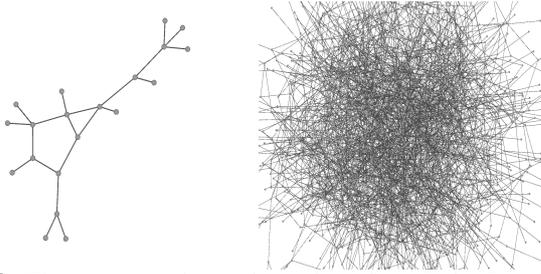

Fig. 2: The non complete sub-graph and an Erdos-Renyi background (1900 nodes, average degree 3).

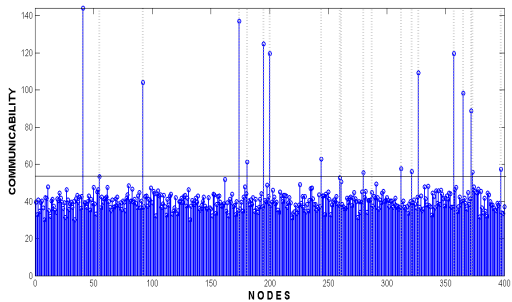

Fig. 3: The selection of candidate nodes. The horizontal line selects the largest 20 *EBTC,* therefore candidates are above the line. The vertical dotted lines indicate the sub-graph nodes. Twenty backgrounds were used to produce this picture.

Table 2: Multiple backgrounds, complete target (average degree 19).

| Background[a] | SW | SW |
|---|---|---|
| nodes | 1024 | 1024 |
| av. deg. bgr. | 40 | 40 |
| nr. backgr. | 2 | 6 |
| % id. rate | 96% | 96.1% |

[a] All backgrounds have density less than 0.08.

*Single background scenario*. The procedure is essentially the same, except that now $N = 1$. In the first experiment the sub-graph is the same of [7], with average degree 2.5. ER and SW backgrounds ranging from 2 to 6 average degree were tested, but the identification rates showed low values, apart from the SW background case that provided a 75% identification rate. Some real graphs were also tested, obtaining meaningful results (45%) only for a local area computer network infected by malware (infected LAN). A bootstrap technique was attempted in order to provide more artificial backgrounds from the one available, but without improvements.

Thus we relax again the problem to the 20 nodes clique. ER, SW and Barabasi - Albert (BA) backgrounds were tested (average degree ranging from 19 to 60, 1024 nodes) achieving high identification rates, see Table 3 for details.

BA results were the worst, not exceeding the 24% identification rate, while SW were the best (96.5%). In fact, BA graphs tend to form many clusters, much more than ER graphs, especially if *n* is not large. In addition, the clusters are inter-linked by the preferential attachment property. Such a topology has high communicability value and consequently the EBTC algorithm undergoes some difficulties to discriminate the target.

Since social networks show both clustering and preferential attachment, EBTC is not an appropriate tool for this field of analysis.

We also tested a protein-protein interaction (PPI), a high tension power grid (PG) and an infected LAN as real backgrounds, obtaining identification rates ranging from 10% (PPI) to 86% (PG) and 93% (infected LAN). This time the bootstrap has produced a significant support to the identification. The poor performance of PPI depends on the spectral gap, as explained in [8].

Hence, the single background gets close to the multiple backgrounds performance relaxing the problem to complete sub-graphs.

Since in this scenario we have only one background, the best case complexity is linear, while the worst case for very large *n* is $O(n \log(n))$.

Table 3: Single background, complete target.

| Background[a] | ER | SW | BA |
|---|---|---|---|
| nodes | 1024 | 1024 | 1024 |
| av. deg. bgr | 39 | 60 | 19.3 |
| nr. backgr. | 1 | 1 | 1 |
| % id. rate | 94% | 96.5 | 24% |

[a] All backgrounds have density less than 0.08.

**Conclusions.** – In this paper we consider the following problem: identify the nodes of a sparse sub-graph (whose size is specified in advance), embedded in a larger (synthetic or real, static or dynamic) background graph, with no *a priori* knowledge of the target or of the background. We have shown that a sparse sub-graph can be recovered from a larger (both in size and average degree) dynamic random graph by means of a deterministic algorithm inspired by SPoG methodologies.

The main tool used is the Estrada-Benzi total communicability, summed trough several realizations of the background graph. The issue of the heavy calculations involved with large adjacency matrices has been reduced dramatically by the Krilov approximation of the matrix exponential. As a consequence, in the worst case the algorithm is linearithmic with the number of nodes.

Increasing the noise (the average degree and the degree of the hubs of the background) over three or four times the average degree of the sparse sub-graph, worsen the performance but relaxing the problem to complete sub-graphs allows again an almost total identification.

In practice generating many realizations of the background may be unfeasible, therefore tests on a single background have been conducted. Simulations show fair results for the sparse sub-graph and very good performances for the complete sub-graph.

This methodology may find many applications in various fields. An example is the temporal extension of centrality parameters and metrics based on temporal shortest path, such as infection immunisation modelling and malware spreading.

However, the generalization to the real world is not immediate, depending on the spectral characteristics of the adjacency matrix and on the preferential attachment rule.

Therefore, the EBTC does not seem suitable to study the social networks, but on the other hand, it is an appropriate choice for the technological networks.

REFERENCES

[1] NEWMAN M. E. J., *EPL,* **103** (2013) 28003.
[2] ABELLO, J., PARDALOS, P. M., RESENDE, M. G. C., Am. Math. Soc., **50** (1999) 119.
[3] TSOURAKAKISAR, C. E., preprint at: rXiv:1405.1477v3 [cs.DS] (2014).
[4] MILLER B., ET AL., *Lincoln Lab. J.,* 20 (2013) 10.
[5] SHUMAN D., ET AL., *IEEE Sig. Proc. Mag.*, **84** (2013) 1053.
[6] Karp R. M., *Reducibility Among Combinatorial Problems*, edited by Miller R. E. and Thatcher J. W., *Complexity of Computer Computations*, New York: Plenum, (1972) p.103.
[7] MILLER B. ET AL., preprint at rXiv:1401.7702v1 [cs.SI] (2014)
[8] BENZI M., and Klimko C., *J. Complex Networks,* **1** (2013) 124.
[9] ESTRADA E. and RODRÍGUEZ-VELÁZQUEZ J. A., *Phys. Rev. E* , **71** (2005) 056103.
[10] AFANASJEW, M., EIERMANN, M., ERNST, O.G. and GÜTTEL, S., *Linear Algebra Appl.,* **429** (2008) 2314.
[11] TANG, J., ET AL., preprint at: rXiv:1305.6974v [physics.soc-ph] (2013).
[12] KHULLER, S., and SAHA, B., *Proceedings of ICALP*, (2009) 608.
[13] Musser, D. R., *Introspective Sorting and Selection Algorithms Software: Practice and Experience*, Vol. 27 (Wiley) 1997, 8 p. 983.